\documentclass[a4paper,fleqn,usenatbib,letter]{mnras}

\usepackage[T1]{fontenc}
\usepackage{ae,aecompl}

\usepackage{graphicx}	
\usepackage{amsmath}	
\usepackage{amssymb}	

\newcommand{\be}{\begin{equation}}
\newcommand{\ee}{\end{equation}} 
\newcommand{\bse}{\begin{subequations}}
\newcommand{\ese}{\end{subequations}} 
\newcommand{\bary}{\begin{eqnarray}}
\newcommand{\eary}{\end{eqnarray}}

\newcommand{\mbh}{ M_{\rm BH}}

\newcommand{\mdotbh}{\dot{M}_{\rm BH}}

\newcommand{\Msun}{\rm M_{\sun}}

\newcommand{\mdotedd}{\dot{M}_{\rm Edd}} 

\newcommand{\K}{\rm K}

\newcommand{\Cvisc}{ C_{\rm visc}}

\newcommand{\Ledd}{\lambda_{\rm Edd}}
\newcommand{\LhX}{L_{\rm HX}}
\newcommand{\Lbol}{L_{\rm bol}}
\newcommand{\ergs}{\mbox{erg}~\mbox{s}^{-1}}

\newcommand{\ckpc}{\rm ckpc}
\newcommand{\pkpc}{\rm pkpc}
\newcommand{\cMpc}{\rm cMpc}

\newcommand{\lsim}{\mathrel{\hbox{\rlap{\lower.55ex\hbox{$\sim$}} \kern-.3em\raise.4ex\hbox{$<$}}}}
\newcommand{\gsim}{\mathrel{\hbox{\rlap{\lower.55ex\hbox{$\sim$}} \kern-.3em\raise.4ex\hbox{$>$}}}}

\newcommand{\REF}{Ref-L100N1504}

\defcitealias{schaye2015}{S15}
\defcitealias{crain2015}{C15}

\pdfoutput=1

\title[Dual AGN in EAGLE]{The abundances and properties of Dual AGN and their host galaxies in the EAGLE simulations}

\author[Rosas-Guevara  et al.]{Yetli M Rosas-Guevara,$^{1,2,3}$\thanks{E-mail: yetli.rosas@dipc.org}
Richard G. Bower,$^{4}$ Stuart McAlpine,$^{4}$
Silvia Bonoli,$^{2,3}$  
\newauthor
Patricia B. Tissera$^{1}$
\\
$^{1}$Departamento de Ciencias F{\'i}sicas, Universidad Andres Bello, Av. Rep{\'u}blica 220, Santiago, Chile.\\
$^{2}$Centro de Estudios de F{\'i}sica del Cosmos de Arag{\'o}n, Plaza San Juan 1, Planta 2, E-44001 Teruel, Spain \\
$^{3}$Donostia International Physics Center (DIPC), Manuel Lardizabal pasealekua 4, 20018 Donostia, Basque Country, Spain \\
$^{4}$ICC, Physics Department, University of Durham, South Road, Durham DH1 3LE, UK \\
}

\date{Accepted XXX. Received YYY; in original form ZZZ}

\pubyear{2015}

\begin{document}
\label{firstpage}
\pagerange{\pageref{firstpage}--\pageref{lastpage}}
\maketitle

\begin{abstract}
We look into the abundance of Dual AGN  in the largest hydrodynamical simulation from the EAGLE project. 
We define a Dual AGN as two active black holes (BHs) with a separation below 30 kpc. We find that only 1 per cent of AGN  with $\LhX\geq 10^{42}\ergs$ 
are part of a Dual AGN system at $z=0.8-1$.
During the evolution of a typical binary BH system, the rapid variability of the hard X-ray luminosity on Myr time-scales severely limits the detectability of Dual AGN. To quantify this effect, we calculate a probability of detection, $t_{\rm on}/t_{\rm 30}$, where $t_{\rm 30}$ is the time in which the two black holes are separated at 
distances below 30 kpc and $t_{\rm on}$, the time that both AGN are visible (e.g. when both AGN have  $\LhX\geq 10^{42}\ergs$) in this period. We find that the average fraction of visible Dual systems is 3 per cent.  The visible Dual AGN distribution as a function of BH separation presents a pronounced peak at  $\sim 20$ kpc that can be understood as a result of the rapid orbital decay of the host galaxies after their first encounter. We also find that $75$ per cent of the host galaxies have recently undergone or are undergoing a merger with stellar mass ratio $\geq 0.1$.  Finally, we find that the fraction of visible  Dual AGN increases with redshift as found in observations.
 
\end{abstract}

\begin{keywords}
galaxies:active --methods: numerical -- quasars: supermassive black holes
\end{keywords}



\section{Introduction}
Supermassive Black Holes (BHs) appear to reside in the centre of all massive galaxies (e.g. \citealt{kormendi2013}).  If galaxy mergers are expected to be common in a hierarchical Universe
\citep{white1978}, then BH mergers should be common too.  
 From numerical studies and models,  we know now that during galaxies mergers, supermassive BHs follow the trajectory of the nucleus of the host galaxies (see the review of \citealt{colpi2014} and all therein references). Subsequently, a supermassive BH binary (at scales of a few parsecs) is thought to be form. The time that will take the BHs to eventually merge will strongly depend on their environment that will be set by the properties of the host galaxies (e.g. \citealt{mayer2007, mayer2013, capelo2015})  such as the  presence of molecular clouds or stellar clusters  (e.g. \citealt{perets2008}),  the effects of the galaxy axisymmetry  and  triaxiality  (e.g. \citealt{khan2013,vasiliev2014,vasiliev2015,gualandris2017}). While it is observationally difficult to study BH binaries, hints on their evolution can be found in the observational properties of BH pairs, defined to be BHs in interactive
 galaxies that have not reached the  binary stage.   Observationally  only active BHs can be easily observed, thus many studies have explored the properties of Dual AGN,  defined as two active BHs  at kpc-scales.


Observational studies suggest that the fraction of Dual AGN is small \citep{fu2011a,rosario2011}.  
\cite{liu2011}  found that the fraction of Dual AGN is $1.3$ percent within a 30 kpc scale using a large study of optical 
AGN pairs at $z<0.16$ with  SDSS  spectroscopy.   However,  detecting Dual AGN at kpc scales in the local Universe is not an easy task since observations at high resolution are needed  \citep{komossa2003,hudson2006, bianchi2008, koss2011a, 
 koss2012, mazzarella2012, shields2012}. For example,  using SDSS spectroscopy could affect the detection of Dual AGN at close scales because of the fibre collision limits. Optical surveys also tend to be incomplete  \citep{hickox2009, koss2011a}. To overcome this difficulty, \cite{koss2012}  select moderate luminous AGN in ultra hard-X-rays along  with optical observations.  They find a much larger Dual AGN fraction: 7.5 percent of their sample are in Dual AGN at a separation of 30 kpcs. The Dual AGN fraction
 goes down to 1.9 percent  when both AGN in the Dual system were only detected using X-ray
spectroscopy.   Another way to find candidates of  Dual AGN is by searching for a double-peaked 
 in the narrow  AGN emission lines  (e.g. \citealt{comerford2009b, comerford2011,barrows2013}). Using this technique along with follow-up observations, \cite{comerford2015}  and  \cite{muller-sanchez2015}   found seven Dual AGN where it was possible to resolve two distinct active nuclei at separations of less than 10 kpcs.   

Whether galaxy mergers enhance AGN activity or not is still under debate. Some of the observational studies mentioned above found that  Dual AGN tend to be in galaxies suffering mergers. This suggests that galaxy mergers enhance AGN activity, at least, for the most luminous AGN (e.g. \citealt{treister2012,donley2018}). For instance, \cite{koss2012} found that the X-ray luminosity of the AGN in Dual systems increases with decreasing galaxy separation, being a galaxy merger the key to activate the AGN. Similarly,  \cite{comerford2015}  found that Dual AGN  in major mergers are more luminous than AGN hosted by no interacting galaxies. On the contrary,  other works (e.g. \citealt{cisternas2011,schawinski2012,hernandez2016,villforth2017} ) find that galaxy mergers do not have a significant role in the AGN activity.  

From the numerical point of view, hydrodynamics simulations of idealised galaxy mergers at high resolution have been used to investigate the Dual AGN activity at the different evolutionary stages of a galaxy merger (e.g. \citealt{blecha2013, vanWassenhove2014, capelo2017}).  For example, \cite*{capelo2017}, based on the work by  \cite*{vanWassenhove2012}, study the importance of the merger conditions and  the properties of the host galaxies on the  Dual AGN.  By varying the initial mass ratio, the gas fraction and the geometries of the merger,  they find the  Dual AGN activity increases after the late pericentric passage. \cite*{blecha2017} also use  hydrodynamics simulations along with dust radiative transfer to explore the mid-IR AGN signatures during the late evolutionary stages of  a galaxy merger.   Although these studies are very insightful  to understand  Dual  AGN activity in galaxy mergers, they  miss the effect of the environment  and  then the occurrence of the Dual AGN in a cosmological context.

The new generation of cosmological hydrodynamics simulations are currently the best tools available to study the incidence of Dual AGN and what drives their activity in a cosmological context.
Previous numerical studies have found Dual AGN  to be rare as well. For instance,  \cite{steinborn2016} use a large simulation with a volume of  (182 Mpc)$^3$  from the suite of Magneticum Pathfinder Simulations,  to compare very close Dual AGN systems to non-active BH pairs and to offset AGN. They define as a BH pair with only one BH active as offset AGN.  \cite{steinborn2016}   found a Dual AGN  fraction to be less than 1 per cent of the total number of AGN at $z=2$.  The non-active BH pairs in their simulation accrete less gas from the intergalactic medium than Dual AGN.   \cite{volonteri2016},  using the horizon-AGN simulation found that the fraction of Dual AGN living in the same host galaxy with a $<30$ kpc separation is 0.1 per cent at $z=0$  for relative massive galaxies independently on whether these galaxies host an AGN or not.  This fraction increases to 2 per cent at $z=1$. They explore the occupation fraction of BHs as a function of stellar mass, finding that the fraction of  Dual systems rises as distances become small.  In the context of Dual AGN evolution, \cite{tremmel2017} follow the evolution of a single Dual AGN (with distances below $1$ kpc) in the most massive halo in the Romulus simulation with a volume of only (8 Mpc$)^3$. They find that this Dual AGN is activated by a major merger.  

An interesting question that arises is the cause of the low frequency of Dual AGN.  It is  because  Dual AGN is an ephemeral phase due to the AGN intrinsic properties such as its variability or it is because of the particular properties of the host galaxies of Dual AGN, such as their stellar mass or their gas fraction or  it  is because of the particular merger history of the host galaxies.  To shed further light on this,  our main goal is to investigate the pure theoretical predictions in the abundance of Dual AGN in X-ray bands for the hydrodynamical cosmological simulation EAGLE.
A series of papers have analysed the galaxy population in  EAGLE finding reasonable agreement with the evolution of the galaxy mass functions \citep{furlong2015}, the evolution of galaxy sizes \citep{furlong2017}, the colour-magnitude diagram \citep{trayford2016}, the properties of molecular and atomic gas \citep{lagos2015,bahe2016}  and 
the oxygen abundance gradients  of the star-forming disc galaxies \citep{tissera2018}. The simulation also reasonably reproduces the evolution of the AGN luminosity functions in X-ray bands up to $z=1$ \citep{rosas-guevara2016} and the different observational trends seen on the plane of star formation and black hole accretion rates \citep{mcAlpine2017}.

In this paper, we explore the abundances and properties of Dual AGN in the largest cosmological hydrodynamical simulation of the EAGLE project \citep{schaye2015,crain2015} at $z=0.8-1$. We also explore the properties of the host galaxies and their recent merger history.
The outline is as follows.  In section \ref{sec:meth}, we describe the simulations and the post-processing analysis to identify Dual AGN.  In section \ref{subsec:movie},  we show the evolution of a Dual AGN as a study case.  In section \ref{subsec:variability}, we explore the effects of AGN variability in the detection of a Dual AGN.
 The Dual AGN fraction as a function of separation is shown in section \ref{sub:distance}. We also investigate the properties of their host galaxies and their recent merger histories in section \ref{sub:hostgalaxy}, and in section \ref{sub:fractionz} we look into the abundances of Dual AGN as a function of redshift.  Finally, in section \ref{sec:discussion} and \ref{sec:summary}, we  discuss and summarise our findings. 
 
\section{Methodology}
\label{sec:meth}
\subsection{ Simulations}
 The Evolution and Assembly of GaLaxies and their Environment (EAGLE, \citealt{schaye2015,crain2015})\footnote{http://www.eaglesim.org}   is a suite of  cosmological hydrodynamical simulations, comprising various galaxy formation sub-grid models, numerical resolutions and  volumes.  
The simulations were performed with a heavily modified  version of SPH code P-Gadget3  \citep{springel2005b} that includes: gas cooling, metal enrichment  and energy input from star formation and black
hole growth.  A full description of the simulation suite can be found in  \cite{schaye2015}, with  the calibration process described in  \cite{crain2015}. 
Here, we concentrate on the largest  simulation with a comoving volume\footnote{We will refer to comoving distances  with a  preceding  `c' ,such as  \ckpc ~, to refer to comoving kiloparsec  and physical lengths will
  be preceded by  a `p'  such as  \pkpc.~} of  $(100 \,\cMpc)^3$, denoted as  \REF.
 The  mass resolution is 
$9.7 \times  10^6 \Msun$ for dark matter particles and $1.81 \times 10^6 \Msun $ for baryonic particles and  a softening length of $2.66\, \ckpc$
   limited to a maximum physical  size of $0.70 \,\pkpc~$.  The simulation adopts the cosmological parameters taken from  \cite*{planck13}.\footnote
  {The values of the cosmological parameters are: $\Omega_\Lambda=0.693$, $\Omega_{\rm m}=0.307$, $\Omega_{\rm b}=0.04825$, $\sigma_8=0.8288$, $h=0.6777$, $n_{s}=0.9611$ and $Y=0.248$.}

The  simulation outputs were analysed using the {\sc subfind} algorithm
to identify bound sub-structures \citep{springel2001,dolag2009} within each dark matter halo. We
identify these substructures as galaxies and measure their stellar
and gas masses within a radius of 30 $\pkpc$ (as per \citealt {schaye2015}).
\footnote{The outputs of the simulation are public available by querying the EAGLE SQL web interface
 http://icc.dur.ac.uk/Eagle/database.php  \citep{mcAlpine2016}} 

\subsection{Black hole sub-grid physics}

Black holes (BH) are seeded into  the minimum potential of  dark matter halos  with masses larger than $1.48 \times10^{10}\Msun $. BHs  
grow via two processes: gas accretion and mergers.  
The  accretion onto BHs is implemented as a modified Bondi prescription, limited to the Eddington rate \citep{schaye2015}.
This modification  modulates the high circulation flows  with a viscous parameter, 
$\Cvisc$,  introduced by \cite{rosas-guevara2015}.  
A fraction of the accreted mass is converted into thermal energy and released stochastically  into the neighbouring gas \citep{booth_schaye2009}. 
The stochastically selected  gas particles around the BH are  heated 
by  a fixed  temperature increment, $\Delta T $,  ($=10^{8.5} \K$ for the  \REF~ simulation).  We highlight that only a single \textit{mode}
of AGN feedback is adopted, independent of the BH mass or halo mass,  using a constant  efficiency of $0.1$, from which, a fraction of $0.15$ 
is coupled to the interstellar medium. 

\subsection{Black hole merger criterion}
\label{sec:mergercriterion}
A full description of the black hole merger criterion can be found in  \cite{booth_schaye2009,schaye2015}. Given its importance towards this study, 
here we provide a brief review.
A BH merger will occur in EAGLE when the BHs: (1) are separated  by a distance below the smoothing kernel
and also below three times the gravitational softening length and  (2) when   the BH relative velocity is 
lower than  the circular velocity  at a separation of $h_{\rm BH}$, $v_{\rm rel} < (G\mbh/h_{\rm BH})^{1/2}$, where
  $h_{\rm BH}$ and  $\mbh$ correspond to the smoothing kernel and the subgrid mass of the most  massive BH of the system.  
 This criterion avoids  a premature BH merger  when their host galaxies are starting to merge. 

Because the simulation can not  adequately model  the dynamical friction  for BHs with masses below the initial mass of the gas,
it is imposed that BHs with $\mbh<100 \,m_{\rm gas}$ are re-located  to the minimum of the gravitational potential of the halo. It is also 
imposed in each step that the BHs change to the position  of the neighbouring  particle with the lowest gravitational potential  of all the neighbouring particles 
with two conditions: (1)  their  velocity relative to the BH is smaller than $0.25c_{\rm s}$ where $c_{\rm s}$  is the sound speed of the local medium surrounding the BHs, and (2)  their distance is smaller than  three gravitational softening lengths.

\subsection{Dual AGN sample}
\label{sec:sample}
 We make use of the 29 snapshots of the simulations that store the full information of the particles between  $z=20$ and $z=0$. Following \cite{rosas-guevara2016}, we  take advantage of the `snipshots', that  are more frequent outputs of the simulation than the snapshots. The snipshots store a  reduced set of the particles properties, but  with a   finer temporal resolution, ranging between $10$ and $60$ Myr. We use  200  of these snipshots in this study.   We also use the log files that  contain properties of the BHs and of their surrounding gas  with a much  better temporal resolution to capture meticulously the evolution of AGN. 

We use the Eddington ratio as a measure of  the activity level of BHs defined the Eddington ratio as  $\Ledd = \mdotbh/\mdotedd$, 
where $\mdotbh$ and $\mdotedd$ are the instantaneous  BH mass accretion rate and Eddington limit respectively. 
The BHs with $\Ledd\geq 10^{-2}$ are considered to be prominent  sources of   luminous X-rays, assuming they are surrounded 
by a thin and efficient nuclear disc and define them as `active'.  BHs with $\Ledd$ below this ratio  and higher than $10^{-4}$ are assumed 
to be enclosed  by a thick and inefficient accretion disc  and they  would not provide a significant contribution of  X-ray luminosity. Finally 
we define  an inactive BH  when  $\Ledd\leq 10^{-4}$.
We note that the threshold value taken  to define  an  `active' BH  does not significantly affect the evolution of the AGN luminosity functions (see \citealt{rosas-guevara2016},  appendix B). 

To remain consistent with  the results of  \cite{rosas-guevara2016} and \cite{mcAlpine2017}, we define  the bolometric luminosity as  10\% of 
the instantaneous mass accretion rate.   The bolometric luminosity is  converted  into  hard X-ray  luminosity (2-10 keV) using the redshift  independent 
bolometric corrections from  \cite{marconi2004}. 

We refer to \textbf{Dual AGN} as  `active' BH pairs with a separation of 30 $\pkpc$ or lower. We exclude  AGN with lower distances than $1 \,\pkpc$ since the 
simulation does not accurately resolve their evolution at such small scales.  We create a sample of \textbf{visible Dual AGN} where both members of the close 
 pair are accreting at  $\LhX\geq10^{42} \ergs$ in a given snipshot. A  \textbf{One AGN} sample is also defined where only one member is above this threshold. With this criterium, 
 at $z=0.8-1$,   there are  109  Dual AGN, 29  of them belong to the visible Dual AGN sample  and 73 to the  One AGN sample. The rest of Dual AGN  are too faint 
 with a hard X-ray luminosity, $\LhX$, between $ 10^{40}\ergs$ and  $10^{42} \ergs$, and therefore are not visible in this band
 even though they could be irradiating near the Eddington limit. To give a sense of the BH masses powering  visible Dual AGN,  the median BH mass is  $4.4\times 10^6 \Msun$ and $70$ percent of the BHs have a  mass larger than $10^6\Msun$. 
 The median   $M_{\rm BH,1}/ M_{\rm BH,2 }$, where $M_{\rm BH,1 }$ and $M_{\rm BH,2}$  are the masses  of  less  and more massive BH respectively, is
 0.1 and  22 percent of the Dual systems  have  a   $M_{\rm BH,1}/M_{\rm BH,2 }$ ratio higher than 0.3.  We do not make any distinction of Dual AGN with
 respect to any property of their host galaxy, except in section \ref{sub:hostgalaxy}.


 
\section{Results}
\label{sec:results}

\subsection{ A study case: The evolution of a Dual AGN system in EAGLE }
\label{subsec:movie}
 \begin{figure}	
	\includegraphics[scale=0.32]{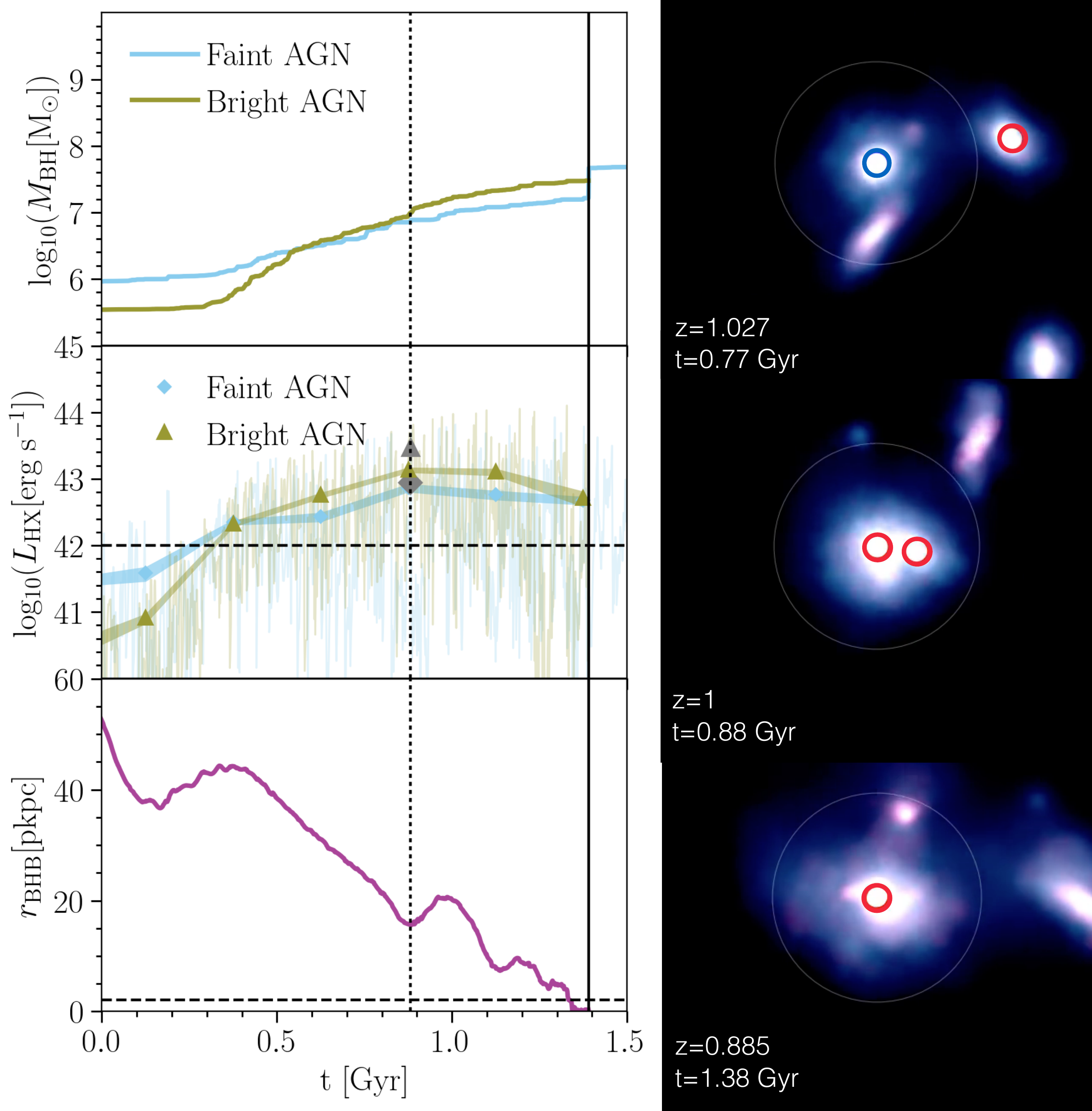}
    \caption{  The evolution of a Dual AGN in the EAGLE simulation. \textit{ Left figure}: The evolution of the BH mass (top plot), 
    of hard-X-ray luminosity (middle plot) and the BH separation  (bottom plot)  in the last 1.4 Gyr before the BHs merged. The markers represent  the mean luminosity over a 250-Myr-period and the filled region the standard error of the mean. The vertical dashed line represents the time at  $z=1$ and 
     the solid line the time they merged. The grey markers represent the $\LhX$ at the moment the Dual AGN was observed ($z=1$). The horizontal dashed line in the bottom panel is the distance at which the merger criterion is applied.  \textit{Right figure}:  Images of the host galaxies  when the BHs are at separations of $ \leq 30\, \pkpc$~at different times as indicated on the labels.    The circles correspond to the positions of the BHs. For each BH, red circles represent  
     $\LhX\geq10^{42}\ergs$  whereas  blue circles the converse. A movie  of  the evolution of the 
     host galaxies can be found in found in footnote 5. The evolution shows that 
     the Dual AGN could be variable at scales of Myrs.}
    \label{fig:movie}
\end{figure}

We begin by  illustrating  the evolution of a typical Dual AGN  observed at $z=1$. In the left panel of Fig.~\ref{fig:movie}, 
we show, from top to bottom, the BH mass, hard X-ray luminosity, and AGN separation as a function of cosmic time relative to the merger, where $t=0$ corresponds to $z=1.3$.
The green  and  light-blue curves represent the evolution of the brighter  and the fainter AGN,  respectively, where the brighter AGN is defined to be so  $z=1$. In the right panel we show  a visualisation of the system at three different times (as labelled). These frames  are part of a movie \footnote{https://www.cefca.es/owncloud/index.php/s/LXh4I0ikwFimR9H} of the host galaxies of the Dual AGN.  

At large separations, the BHs have masses within an order of magnitude of the seed mass ($1.48 \times10^{5}\Msun$). Both BHs then gradually acquire 
mass via gas accretion and  they sporadically vary between  the states that we defined as `Dual AGN', `One AGN' and an `inactive pair'. Note that the average $\LhX$ for both BHs is above $10^{42}\ergs$   (see the horizontal dashed line) just after 400 Myrs. At this stage, they have comparable 
BH masses of $6.3\times 10^{6}\Msun$  and  $10^7\Msun$ at $z=1$ (see the vertical dashed line) in their Dual AGN phase. This is also where the BHs are located  in the pericentre of their orbit (see bottom panel).  

The BH pair continues to grow by gas accretion funnelled by  the galaxy encounter and eventually merge to form a final BH with a mass of $2.7\times 10^{7}\Msun$. 
 Interestingly, the luminosities in the hard X-ray band vary by two orders of magnitude over a temporal scale of megayears 
 as seen in the middle plot. However, overall, the luminosity increases as the BHs get bigger and as their host galaxies get closer to each 
 other (see markers in the middle plot). In the images,  the circles represent the location of each AGN, coloured blue  if the AGN is too faint to be 
 visible in the hard X-ray band and red when it is visible.  The figure shows that the luminosity of both AGN increases on average as the distance between the host galaxies reduces, but also 
  that the presence of rapid AGN variability will significantly reduces the detectability of Dual AGN. 
\subsection{The Effects of AGN variability}
\label{subsec:variability}
As  Fig.~\ref{fig:movie}  has shown, the variability in the AGN luminosity can affect the detection of Dual AGN. 
To quantify the significance of this effect, we  measure $t_{\rm on}$/$t_{30}$, where $t_{30}$ is the time   the BH pair 
spend with a separation  $\leq 30\, \pkpc$ and $t_{\rm on}$ is the time that  the BH pair is `turned on' during this period (i.e., when both  
AGN are accreting at  $\LhX\geq10^{42}\ergs$). This ratio,  $t_{\rm on}$/$t_{30}$, is a proxy for 
the probability of detecting  a Dual AGN: If this ratio is 1,  the BH pair is always `turned on'. When this ratio is 0 implies that it is impossible 
to detect the BH pair at these distances.  

\begin{figure}	
\includegraphics[width=1.\columnwidth]{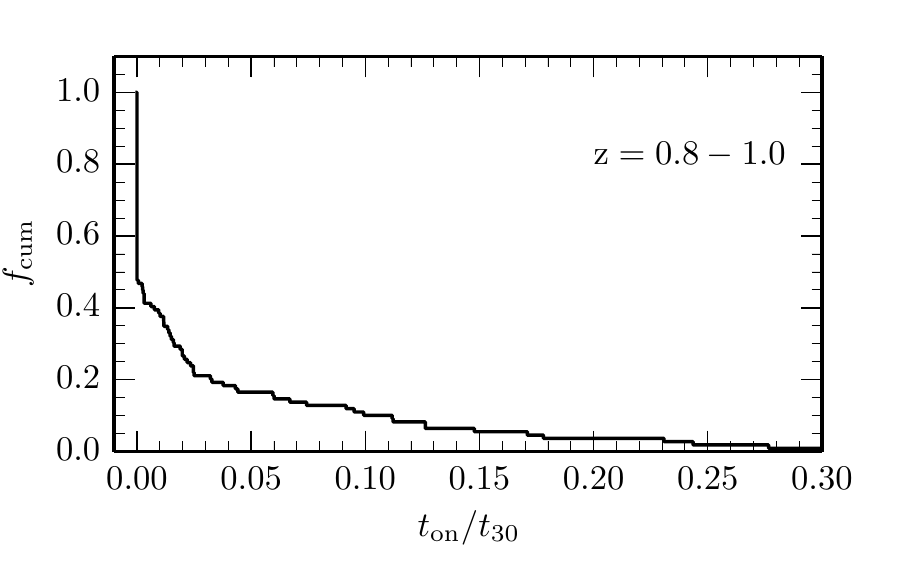} 
    \caption{The cumulative Dual AGN fraction  as a function of  $t_{\rm on}/t_{30}$  at $z=0.8-1$, where $t_{\rm on}/t_{30}$ is the fraction of the time that both AGN in 
    a Dual system are `turned on' during the time  they are separated by less than $30\, \pkpc$. This quantity  act as a proxy for the detection 
    probability of a Dual AGN. The maximum detection probability that a Dual AGN present is  0.4 and  $40$ percent ofDual  AGN  have a detection probability larger than 0.01.}
\label{fig:ratiotonttot}
\end{figure}

Fig.~\ref{fig:ratiotonttot} shows the cumulative Dual AGN fraction as a function  of  $t_{\rm on}/t_{30}$  at $z=0.8-1$. 
The figure illustrates that 60 percent of  Dual AGN have a probability of being detected smaller than 0.01 and only 10 percent 
of Dual AGN are `turned on'  for more than 10 percent of the time. With this result, we can estimate the average  number  of visible Dual AGN  respect to  the total number of 
Dual AGN. We found that this average fraction is similar to 3 percent, meaning that from 100 Dual AGN in a given volume, only 3 of them are going to have a  hard X-ray luminosity above  $10^{42}\ergs$. 





\subsection{Incidence of Dual AGN as a function of separation}
\label{sub:distance}
\begin{figure}	
	\begin{tabular}{c}
	\includegraphics[width=1.\columnwidth]{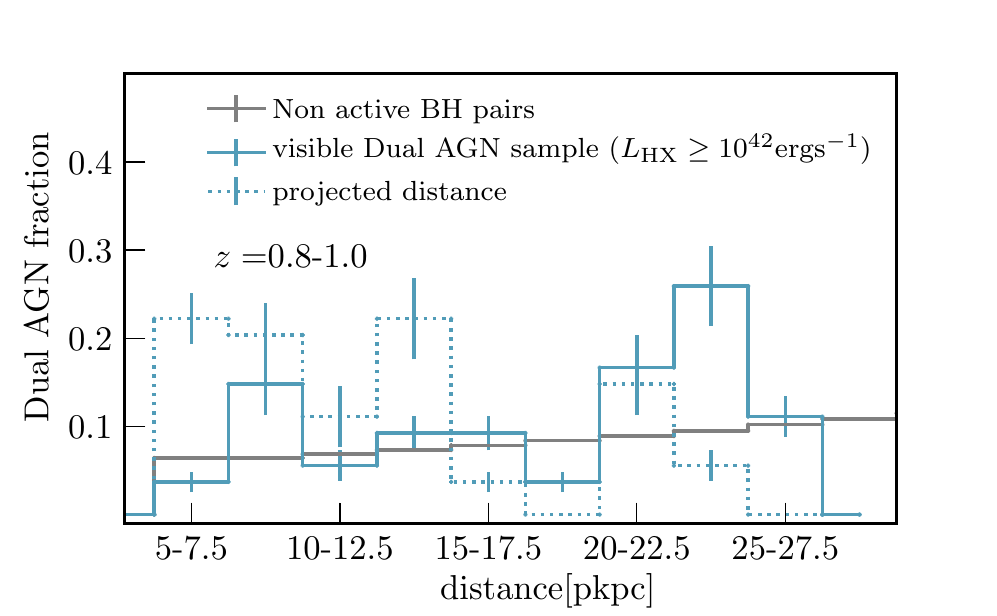} \\
	\includegraphics[width=1.\columnwidth]{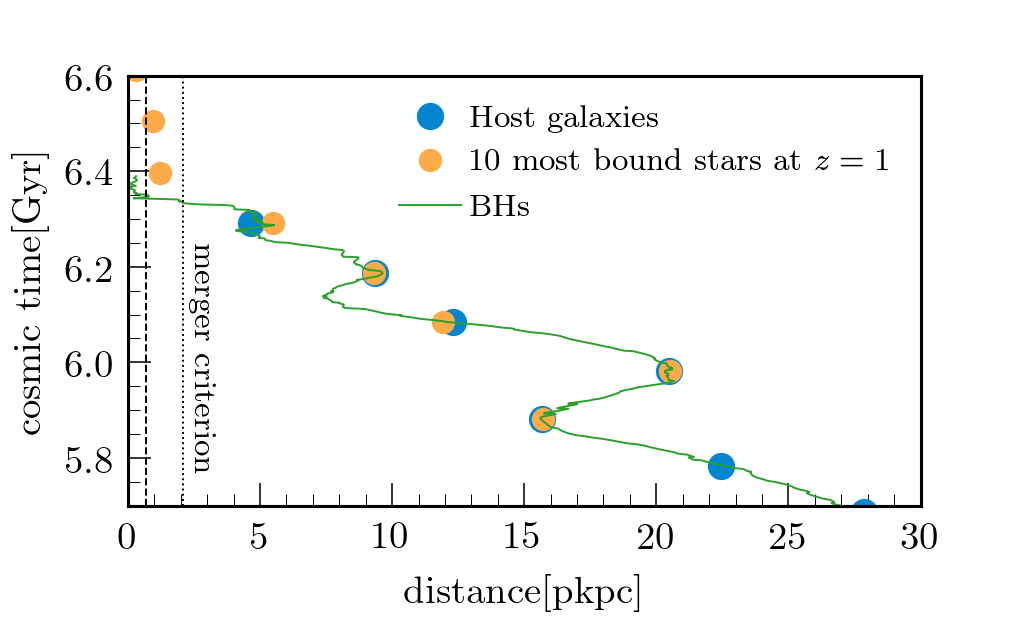}
	\end{tabular}
    \caption{ \textit{Top panel:} The normalised distribution of visible Dual AGN  as a function  of the separation  (blue solid line) and projected  separation (blue dotted line) at $z=0.8-1$.  
    Bars correspond to the  standard errors of the mean using the outputs of the simulation in this redshift bin.  We excluded bins below $5 \,\pkpc$ to avoid sensitivity to the 
    merge criterion in EAGLE.  The distribution presents a  peak at $\sim 20-25\,\pkpc$.  
    By contrast, the fraction of inactive BH pairs (grey line) increases with increasing separation. \textit{Bottom panel:}
      A typical example of the cosmic time as a function of separation of the BHs (green line), the centre of the potential of the  host galaxies (blue circles) and the ten most bound star particles (orange circles) of each galaxy at $z=1$. The plot highlights the rapid evolution of the host galaxies of a typical Dual AGN  after their last encounter and the small effect of the reposition of the BH.}
    \label{fig:frac_dist}
\end{figure}

One of the observational features of Dual AGN that is frequently invoked is their increasing incidence as a function of separation. We 
investigate  this feature by using the visible Dual AGN sample defined in section \ref{sec:sample}. 

The top panel of Fig.~\ref{fig:frac_dist} shows the predicted average fraction of  visible Dual AGN as a function of the separation between BHs at 
$z=0.8-1.0$ (blue solid line). We do not consider Dual AGN with separations smaller than $5\,\pkpc$ since their behaviour could be affected  by  the BH merger criterion  (see \ref{sec:mergercriterion})\footnote{Because the maximum  proper softening length is 0.7 pkpc, the merger will take place when the BHs  have a distance below 2.1 pkpc}.  At small distances ($<15\,\pkpc$), there is no clear trend in the distribution while the Dual AGN fraction presents a pronounced peak  located around $20$ and $25\, \pkpc$.  The presence of  the peak is independent of  the binsize  and  the  redshift (not shown here).   
The peak  also remains when taking  projected distances  but is  shifted to smaller separations as shown in the top panel of  Fig.~\ref{fig:frac_dist} (blue dotted line).

To understand its origin, we follow the separation of the BHs and of their host galaxies through time.   
The bottom panel of Fig.~\ref{fig:frac_dist} shows that the BHs (green curve) and their host galaxies (blue circles) follow similar trajectories before the galaxy 
merger takes place. The last close encounter, occurs at $20-25\, \pkpc$ and then the host galaxies 
rapidly spiral inwards, merging at $5-10\,\pkpc$.  This last encounter between galaxies could drive gas towards the central parts of the galaxies and  feeds the BHs creating a Dual AGN (see section \ref{sec:discussion}).
It takes much longer for the BHs  to eventually merge, spending more time at closer separations.  

The shape of the distribution  is  also contrasted with the distribution of  non active BH pairs  (grey line), defining  them as any 
two BHs whose Eddington ratio is $\leq 10^{-4}$.  The fraction of non active BH pairs gradually rises with larger distances contrasting  the behaviour of the visible Dual  AGN fraction. 

Something to note is that the Dual AGN fraction may be affected by the reposition of the BHs to the minimum of the  potential of the host halo (see \ref{sec:mergercriterion}). To investigate the possible effect of the BH reposition,  we follow the trajectories of 
the 10  most bound star particles of each host halo  at $z=1$ and compare to the trajectories of the BHs. Since the star particles do not undergo any relocation, if the effect of BH repositioning were small, the trajectories of the BHs and  of the 10 most bound star particles would be similar.  Indeed,  we find that in most cases  the trajectory of the BHs and of the 10 most bound star particles are almost identical for distances above $5\,\pkpc$. A typical example  of this is shown in  the bottom panel of Fig.~\ref{fig:frac_dist}.  Therefore the BH relocation does not significantly  affect  the Dual AGN fraction. Moreover, the BH relocation technique seems to satisfactorily reproduce the BH orbits at kpc-scales.




\subsection{Properties of the host galaxies of Dual AGN}
\label{sub:hostgalaxy}
In this section, we  explore  the main properties  and the recent merger history  of the galaxies that host visible Dual AGN  at $z=0.8-1$. 
Firstly, we look at whether the host galaxy/galaxies  underwent in the last 2 Gyrs  or are undergoing a merger with stellar mass ratio,$f_{M_{*}}$, larger than 0.1. 
Table \ref{table1} summarises the recent merger histories of the host galaxies: $30$ percent of the 
visible Dual AGN reside in the same galaxy whereas the remainder in distinct host galaxies that  are currently interacting.  
From  the subsample of visible Dual AGN living in the same host galaxy (30 percent of the total sample),  60 percent of their host galaxies experienced  a merger with $f_{M_{*}}\geq 0.1$ in the last 2 Gyr.   Fo the subsample of  Dual AGN in distinct galaxies  (70 percent of the total sample)  the fraction is  $0.81$.  In total, 75 percent of the Dual AGN host galaxies had a major merger in their recent history. 

\begin{table}
\caption{The fraction of visible Dual AGN that each resides in a single host galaxy, or two distinct host galaxies, and the fraction of these galaxy subsets that have undergone (or are undergoing)  a merger with $f_{M_{*}}\geq 0.1$ in 2 Gyrs. As an example, if we have 100 visible Dual AGN, in average, 70 of them will be i
n distinct galaxies from which 57 are undergoing a significant merger.}
\begin{tabular}{|l|l|l|}
\hline
                                       &        total fraction    &     undergoing \\
                                       &                                &  /underwent merger\\
                                       &					&      with $f_{M_{*}}\geq 0.1$ in 2 Gyr \\
Dual AGN  in 1 gal         &             0.30            &                       0.60                                    \\
Dual AGN  in 2 gal &       0.70           &                              0.81  \\       
\hline
\end{tabular}
\label{table1}
\end{table}

\begin{figure}	
\begin{tabular}{c}
\includegraphics[width=1.\columnwidth]{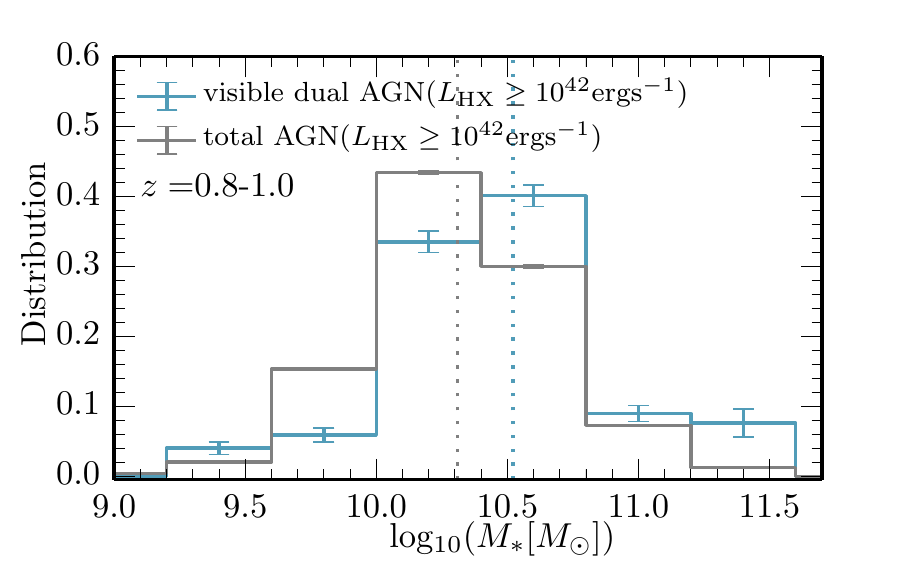} \\
\includegraphics[width=1.\columnwidth]{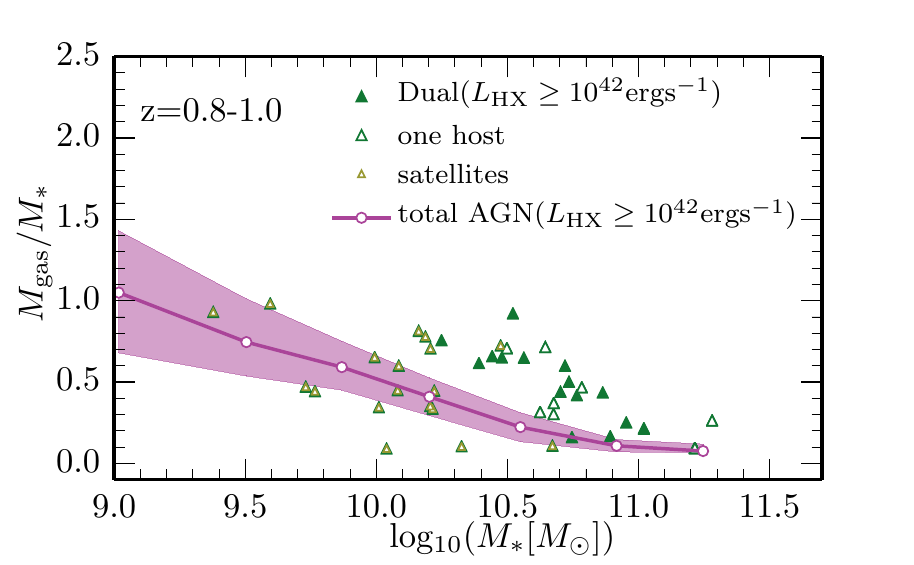}
\end{tabular}	
    \caption{ \textit{Top panel}: The  stellar mass distribution of the AGN host galaxies at $z=0.8-1$. The blue solid line represents the average fraction of galaxies hosting a visible Dual AGN, the grey solid line the galaxies hosting  one visible AGN. Bars correspond to the standard error of the media for all the outputs of the simulation in this redshift bin. The error of the mean of the total AGN population  is  small and not visible. The vertical dotted lines  correspond to the median of each population.  The plot highlights that visible Dual AGN are relative common in massive galaxies.  \textit{Bottom panel}: The gas to stellar mass ratio as a function of stellar mass at $z=0.8-1.0$. 
    The purple solid line with circles represents the median gas to stellar mass ratio in the host galaxies with a visible AGN. The coloured region represents the 30$^{\rm th}$ and 70$^{\rm th}$ percentiles of the distribution.The dark green open triangles correspond to the individual galaxies hosting a  visible Dual AGN, the dark green filled triangles to the central  galaxy of the Dual systems living in interacting galaxies and the light--green open triangles to those that are satellites. Most of the Dual AGN systems  live in a  galaxy with an unusually high gas to stellar mass ratio,  apart from poor gas satellites  that are interacting with their central galaxy.}
\label{fig:hostgalaxy}

\end{figure}

 We have also  investigated the distribution of  stellar masses of the host galaxies of visible Dual AGN  and compare to 
the stellar mass distribution of the galaxies  hosting  at least one AGN with $\LhX>10^{42}\ergs$.  Visible Dual AGN (blue solid line of Fig.~\ref{fig:hostgalaxy}, top panel) 
tend to live in more massive galaxies in comparison with the host galaxies of the  full AGN population (grey solid line). 
The median stellar mass  is $\sim 10^{10.5}\Msun$ (blue dotted line)  0.2 dex higher than that of the total AGN population (grey dotted line). 
Finally, we investigate  the gas to stellar mass  fractions  in the bottom panel of Fig.~\ref{fig:hostgalaxy}. 
We compare the median gas to stellar mass fractions (purple solid line with circles) to the one of the host galaxies of visible Dual AGN (green triangles).
We find that the galaxies hosting a Dual AGN present higher  gas to stellar mass fractions than the median of the distribution, apart from  a few cases.
These few cases (light green open triangles) are satellites that could have run out of gas because of the interaction with the central galaxy. 



\subsection{ The evolution of  the Dual AGN fraction with redshift}
\label{sub:fractionz}
\begin{figure}
	\includegraphics[width=1\columnwidth]{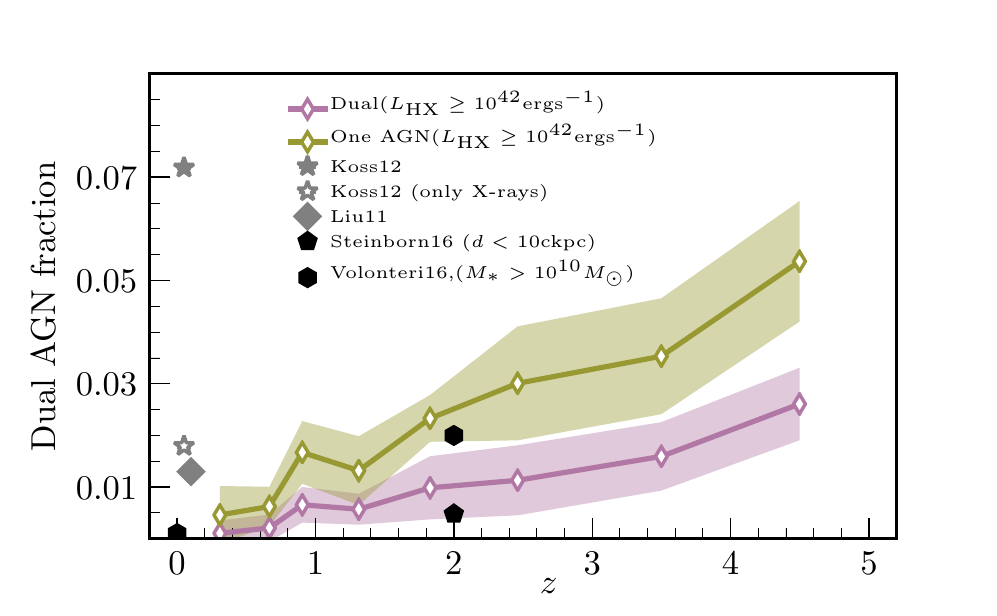}
    \caption{The fraction of visible Dual AGN (purple solid line) and One AGN (green line) as a function of redshift.  The coloured regions correspond to the standard deviation.
    Observational estimates are presented as grey stars (\citealt{koss2012}, not filled grey stars correspond to detection only in hard X-ray bands) and as
    grey diamonds  \citep{liu2011}. Numerical estimates as black  pentagons  \citep{steinborn2016} and hexagons   \citep{volonteri2016} . The plot illustrates
     an increasing trend  in the Dual AGN fraction with redshift and it is much lower compared to observations of the local Universe.}
     \label{fig:AGNfrac-redshift}
\end{figure}

Fig.~\ref{fig:AGNfrac-redshift} shows that the average fraction of visible Dual AGN (purple solid line). Here, we refer the fraction of visible Dual AGN  as the number of AGN that belongs to
the visible Dual AGN sample over the total population of AGN with $\LhX>10^{42}\ergs$.  This definition is similar as observational estimates calculate. 
This fraction increases with redshift, from  $0.1$ percent  at $z=0.0-0.5$  to  $3$ percent at $z=4-5$.
If one of the AGN of the Dual system is only visible in the hard X-ray band (i.e. the One AGN sample),  the  average fraction of Dual AGN  increases for a given redshift as  the 
green line shows. For instance  at $z=0.8-1.0$, it increases from $1$ to $2$ percent.  Note, however, that the average always remains small at all redshifts($\lsim 5\%$).

We perform a qualitative comparison of our results with the current observations in the local Universe.  \cite{koss2012}(empty stars) 
combine X-ray and optical observations to detect close Dual AGN (distance $\leq 30\, \pkpc$), finding  a Dual AGN fraction of $7.5$ percent. 
The Dual AGN  sample of \cite{koss2012}  detected with X-ray spectroscopy and not with emission lines diagnostics, 
decreases to $2$ percent  (filled stars).  \citealt{liu2011}  (diamonds)  use a sample from the 
Seventh Data Release of the SDSS survey at $z=0.1$  based on diagnostic emission line ratios.  They estimate a  
Dual AGN fraction with separations $\leq 30 \,\pkpc$, to be $1.3$ percent. These fractions are marginally above 
the simulation prediction. This could be due to  the different selections and methods used to find  Dual AGN in these studies whose
estimates are also discrepant to each other by similar margins. Nonetheless,  in both  observations and the simulation, Dual AGN are rare.


We additionally include estimates from other cosmological hydrodynamical simulations. \cite{steinborn2016} estimate the fraction 
of very close Dual AGN  (separations of  $\leq 10\, \ckpc$) at $z=2$ from a simulation that is part of the  Magneticum Pathfinder set. This simulation has  a similar  
resolution to that from EAGLE with a larger volume, but only run to $z=2$. They consider an AGN to be a  BH powering at $\Lbol\geq10^{43}\ergs$ ($\LhX\gsim  10^{42}\ergs$). They 
found  a Dual AGN fraction of $0.5\%$  (pentagon), that is below our prediction but it is consistent  because they  only consider  very close Dual AGN ($< 0.33\pkpc$) .  \cite{volonteri2016} 
also estimate a Dual AGN fraction (hexagons)  $z=0$ and $z=2$, in the cosmological hydrodynamic simulation Horizon-AGN. The volume and resolution of the simulation  are similar to EAGLE, but they calculate the Dual AGN fraction differently.  They calculate  the number of  AGN with  $\Lbol\geq10^{43}\ergs$  and hosted by a single galaxy with stellar mass  $\geq 10^{10}\Msun$ over the whole population of galaxies above this stellar mass threshold, independently if they host an AGN or not. 
The discrepancy between the predictions of EAGLE and Horizon-AGN could be due to the different  definition of  the Dual AGN fraction. Beside this,  they also find an increasing  trend with redshift.

\section{Discussion}
\label{sec:discussion}
\begin{figure}
	\includegraphics[width=1\columnwidth]{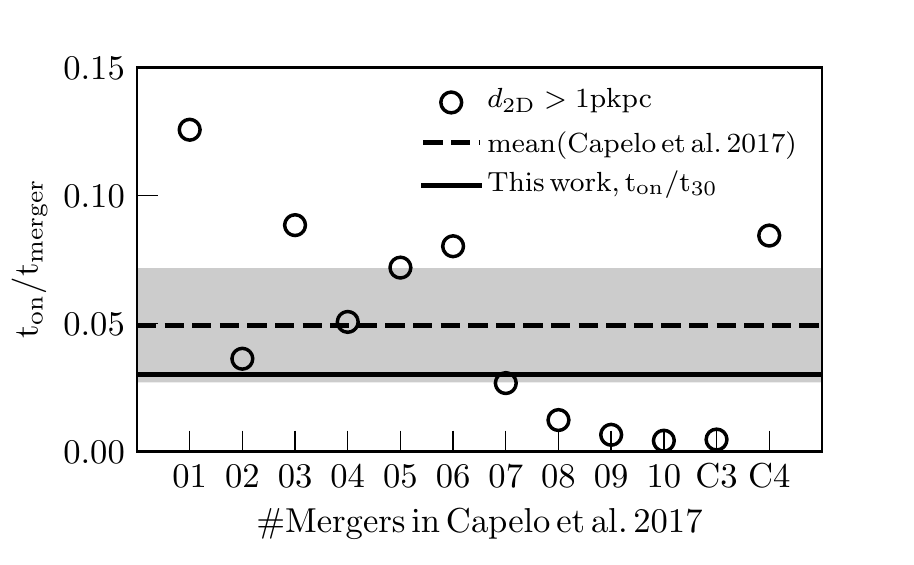}
    \caption{The dual-activity time of the 12 simulations of idealised galaxy mergers from \citealt{capelo2017} normalized by the merger time (initially at a  separation $90 \pkpc$) at projected distances larger than $1\,\pkpc$ and with bolometric luminosity larger than $10^{43}\ergs$. The dashed line corresponds to the median and the shaded region to the 2-sigma error of the median. The solid line corresponds to the probability of detect a Dual AGN in this work. This is similar to the dual-activity time of our Dual AGN but normalised for the time the BHs spend in a separation of $30\,\pkpc$ or lower. We find that our estimate is roughly consistent with this work.}
     \label{fig:comp}
\end{figure}

In this section, we compare our results to high-resolution hydrodynamical simulations of idealised galaxy mergers (e.g. \citealt{vanWassenhove2012, capelo2017,blecha2017}). We discuss the possible effects of the subgrid physics on the results. Finally,  we briefly discuss the main mechanisms that could transport gas in the central part of the galaxies during the merger.
\subsection{Comparison to  other works using simulations of idealised galaxy mergers}

The goal of idealised galaxy mergers simulations is to study the effects of mergers on the galaxy properties at small scales.  These simulations have been also used to study the activity of Dual AGN during a merger. For instance,  \cite*{blecha2017} study AGN activity during a major merger by mimicking mid-infrared WISE observations. They determine the possible factors that could affect the detection of a Dual AGN. The results of these studies are complementary to our work since cosmological simulations operate on larger scales.
 
Recently, \cite*{capelo2017} perform a suite of 12 high-resolution simulations of idealised galaxy mergers,  similar to the setup of  \cite{vanWassenhove2012}. In their study, the galaxies have initial mass ratios larger than 0.1 and disc gas mass fractions of 30 and 60 per cent with different initial geometries of the galaxy encounter (see their Table 1 for the details of each simulation).  They define dual-activity time as the time that both AGN are turned on during the merger, normalised by the duration of the merger. They consider that the merger starts when the BH pair is separated by 90 $\pkpc$. 
This relates to our probability of detection, however, we normalise by the time when the BH pair is first separated by 30 $\pkpc$ (see section \ref{subsec:variability}). The difference
in the definitions can cause some discrepancies in our results.
 Fig. ~\ref{fig:comp} shows the dual activity time at  projected distances larger than 1 $\pkpc$ (just above our resolution limit) and AGN bolometric luminosities larger than $10^{43} \ergs$ (similar to our cut in $\LhX$).  As Fig.~\ref{fig:comp} shows,   our estimate is consistent with the results from \cite*{capelo2017} within $2\sigma$ errors (shaded region).   
The small discrepancy between studies could be because our visible Dual  AGN sample includes major and minor mergers while their simulations only include major mergers.  Also, it is possible that the predictions of the EAGLE simulation underestimate the Dual AGN activity below its spatial resolution. 
 
In agreement with our results, they find that the dual-activity time increases as the BH separations reduce. During the evolution of the merger, there is an increase in AGN activity after the last pericentric stage, agreeable with the peak in the Dual AGN distribution shown in Fig.~\ref{fig:frac_dist}. They also find that the gas fraction has an impact on the Dual AGN activity. This is consistent with our findings in section \ref{sub:hostgalaxy}, where the primary host galaxy of Dual AGN tends to have higher gas mass fractions than the median of the total galaxy population hosting an AGN. Overall, we are in agreement with most of their findings.

\subsection{The effects of the BH merger criterion and resolution} 
As has been previously mentioned, the distribution of Dual AGN as a function of the BH separation could be affected at small distances because of the 
BH merger criterion (see \ref{sub:distance}). We investigate its effects on the results of the paper. 
We obtain similar findings when we exclude Dual AGN at separations below $2.1\, \pkpc$(below this value, the merger criterium could be applied). 
The population of Dual AGN living in the same host galaxy are dominated by  Dual systems with a separation below $2.1\, \pkpc$.  However, 
our main result is preserved since major mergers are also found in the majority of Dual AGN systems.  
The average fraction of Dual AGN detected is not highly sensitive to the merger criterion. This is because the median fraction of the time that Dual AGN are turned on has a low dependence on the BH separation. This is also present in high-resolution idealised galaxy merger simulations \citep{capelo2017}  for Dual AGN at low bolometric luminosities.  Finally, the increasing trend found in the cosmic evolution of the Dual AGN fraction is preserved, however, the Dual AGN fractions tend to be marginally smaller.

\subsection{Other caveats}

Something to discuss is the sensitivity of the prediction of EAGLE simulations due to the subgrid physics, particularly, due to  AGN and star formation feedback. 
\cite*{crain2015}, in an extensive study, prove that the subgrid physics affect the properties of the galaxies in EAGLE. In their study, they use simulations with a comoving volume of $(25\cMpc)^3$  which is far too small for the given frequency of Dual AGN. Although, 
\cite{rosas-guevara2016} compare the evolution of the AGN luminosity functions in hard X-ray bands for three simulations with different subgrid models of star formation and AGN feedback, finding good agreement between them. Therefore, we expect that the Dual  AGN frequency is not significantly affected by the subgrid physics in a cosmological context. Nonetheless, it is something that needs further investigation with improved resolution and larger volumes.

One of our primary results is that galaxy mergers may be a prominent triggering mechanism for Dual AGN activity (see Table 1). 
This leads us to briefly summarise the possible physical processes that transport gas to the centre of the galaxies during a merger.  Using high-resolution idealised galaxies mergers,  \cite{blumenthal2018} explore three pure mechanisms (excluding star formation and AGN feedback) during a galaxy merger. These are (1) clump driven inflow (e.g. \citealt{duc2004}); (2) ram pressure sweeping (e.g. \citealt{capelodotti2017}) and (3) mode driven inflow (e.g. \citealt{barnes1991}). They find that the nature of the resulting inflow depends on the geometry of the encounter.  They suggest that the main mechanisms could be the clump driven inflow. In this process, shock fronts form after the first pericentric distance and form gas filaments that become Jeans-unstable,  forming massive and dense clumps. However, the formation and evolution of these massive and dense clumps could be affected when star formation and AGN feedback are included (e.g. \citealt{hopkins2013}). 

 To test these scenarios in a cosmological context, it would require a more thorough investigation because of the low resolution and the limited volume in the simulations of the EAGLE project. 
We will reserve this study on a future project that will involve the next generation of hydrodynamical simulations with improved resolution and more realistic physical implementation.

\section{Summary}
\label{sec:summary}
In this paper, we investigate the abundance and evolution of Dual AGN in the  EAGLE simulation. We select Dual AGN   as black holes (BH) 
 with Eddington ratios $\geq 0.01$, visible in the hard-X-ray band ($\LhX\gsim  10^{42}\ergs$) and with a separation less than $30\,\ckpc$.  
 We also explore the main properties of their host galaxies and their recent merger histories.   Our main results are as follows: 
\begin{itemize}
\item We show a study case of the evolution of a typical Dual AGN in the simulation at $z=1$ in the last 1.4 Gyrs before the BHs merge. We find that AGN activity is, overall,  
triggered by gas funnelled by mergers of their host galaxies. We also find that rapid variability in hard X-ray luminosities at scales of Myrs is present (Fig.~\ref{fig:movie}). 

\item We explore the effects of AGN  variability on the detection of a Dual AGN at $z=0.8-1.0$. We define a probability of detection, $t_{\rm on}/t_{30}$,  where $t_{30}$ is the time that  BHs 
spend within a separation  $\leq 30 \pkpc$ and $t_{\rm on}$  the time that both AGN are `turned on' during this period (i.e., when both  
AGN have  $\LhX\geq10^{42}\ergs$).  We find that 60 percent of our Dual AGN sample have a probability of detection below $1$ percent (Fig.~\ref{fig:ratiotonttot}). 

\item We obtain the Dual AGN fraction as a function of their BH separation at $z=0.8-1.0$ (Fig.~\ref{fig:frac_dist}). The distribution presents a peak around  20 $\pkpc$. This  could be attributed to the fast evolution of the host galaxies after their last encounter,  which occurs  around this distance.   We find  similar results  for different redshifts and using  projected distances.
This could be ascribed to the gas of the central galaxy being accreted by the infall BH in agreement with observations \citep{lambas2003,alonso2007}.  
 
\item We explore the main properties of the host galaxies and their recent merger histories. We find that 75 percent of the host galaxies recently undergone or are undergoing a merger with a mass ratio larger than 0.1.  We compare the properties of the host galaxies  to the total galaxy population hosting an AGN. We find that  Dual AGN tend to live in galaxies with higher stellar mass and higher gas to stellar mass fractions. We also find that some of the host galaxies of Dual AGN  have a lower gas to stellar mass fraction, but those are satellite galaxies whose BH could be feeding by gas from the central galaxy (Fig.~\ref{fig:hostgalaxy}).  

\item The average visible Dual AGN fraction  in hard X-ray ($\LhX>10^{42}\ergs$)   increases with  redshift (Fig.~\ref{fig:AGNfrac-redshift}).
This rising trend is also found in other  numerical and observational works  \citep{comerford2014, volonteri2016}. When only one of the AGN has $\LhX\geq10^{42}\ergs$, the average Dual AGN fraction increases.  
\end{itemize}


This paper uses a state-of-the-art cosmological simulation to study the evolution of galaxies. The EAGLE simulation  reproduces the observables of galaxies in the local Universe such as stellar masses, colours, sizes. It also reproduces with good agreement the evolution of AGN luminosity functions up to $z=1$ and the contrasting observed trends of the plane of star formation rate--black hole accretion rate. In this paper, the EAGLE simulation allows us to study in a more statistical mean the frequency of  Dual AGN and the effects of  AGN variability in the detection of Dual AGN at $z=0.8-1$.  Dual AGN tend to be rare  and their detection are affected by AGN variability. 
 The enhancement in the fraction of Dual AGN at small scales  is a natural result of the evolution of their host galaxies merging. 
 It also confirms earlier suggestions that Dual AGN might be triggered by significant galaxy mergers. Although the evolution of the Dual AGN is not  captured at scales below the  BH merger criterion is applied in EAGLE, we show that our findings are preserved. Something that it is not completely clear from our work is the conditions of the major mergers to activate a Dual AGN or if major mergers always activate a Dual AGN.  In further work, we plan to extend this study  in more detail.




\section*{Acknowledgements}
We thank Joop Schaye, Lisa Steinborn, Sergio Contreras and  David Izquierdo-Villalba  for the useful comments in the first draft  and  the anonymous referee
for constructive feedback that improved the paper. This work was financially supported by Fondecyt 1150334 Conicyt and from the European Union Horizon 2020 Research and Innovation Programme under the Marie Skodowska-Curie grant agreement No 734374.
RGB acknowledges the Science and Technology Facilities Council (STFC) through grant ST/P000541/1 for support. This work would have not be possible without Lydia Heck's technical support.
We acknowledge the Virgo Consortium for making their simulation data available. The EAGLE simulations were performed using the DiRAC-2 facility at Durham, managed by the ICC, and the PRACE facility Curie based in France at TGCC, CEA, Bruy�res-le-Ch�tel.
This work used the DiRAC Data Centric system at Durham University, operated by the Institute for Computational Cosmology on behalf of the STFC DiRAC HPC Facility (\url{www.dirac.ac.uk}). This equipment was funded by BIS National E-infrastructure capital grant ST/K00042X/1, STFC capital grant ST/H008519/1, and STFC DiRAC Operations grant ST/K003267/1 and Durham University. DiRAC is part of the National E-Infrastructure.  We thank contributors to SciPy \footnote{http://www.scipy.org} , 
Matplotlib \footnote{http://www.matplotlib.sourceforge.net} , and 
the Python programming language \footnote{http://www.python.org}

\bsp	
\label{lastpage}
\end{document}